\documentclass[aps,prb,twocolumn,groupedaddress]{revtex4}

\usepackage{graphicx}
\usepackage{dcolumn}

\begin{document}


\title{Ferromagnetism in Fe-doped Ba$_{6}$Ge$_{25}$ Chiral Clathrate}


\author{Yang Li}
\author{Joseph H. Ross, Jr.}
\affiliation{Department of Physics, Texas A\&M University, College Station, TX 77843-4242}


\date{\today}

\begin{abstract}
We have successfully synthesized a Ba$_{6}$Ge$_{25}$ clathrate, 
substituting 3 Fe per formula unit by Ge. This chiral clathrate has Ge sites 
forming a framework of closed cages 
and helical tunnel networks. Fe atoms randomly occupy these sites, and exhibit 
high-spin magnetic moments.  A ferromagnetic transition is observed with 
$T_{c}$ = 170 K, 
the highest observed $T_{c}$ for a magnetic clathrate.  However, the magnetic 
phase is significantly disordered, and exhibits a transformation to a re-entrant 
spin glass phase. This system has a number of features in common with other dilute 
magnetic semiconductors.
\end{abstract}


\maketitle


Si, Ge and Sn clathrates have received increasing attention
over the past few years. These materials have a 
semiconducting framework into which metal atoms can be substituted, 
providing a number of possibilities for new materials,\cite{roy92} for example 
new thermoelectrics,\cite{nolas98,sales01} and
superconductors.\cite{kawaji95} Ferromagnetism has been observed 
in Mn-doped Ge clathrate, with $T_{c}$ = 10 K,\cite{kawaguchiapl00} 
and clathrates having Eu encapsulated in the cages, with $T_{c}$ up to 
37 K.\cite{sales01} Currently there is strong interest in ferromagnetic 
semiconductors for spintronic applications,\cite{ohno00} and much 
effort has been devoted to the magnetic doping of semiconductors
using low-temperature epitaxy.\cite{ohno03} Clathrates 
offer the possibility of semiconducting phases containing 
magnetic atoms at equilibrium, which might 
be compatible with conventional substrates.\cite{munetoh01}
Here, we report the synthesis of a stable Fe-doped Ge 
clathrate, and the observation of a ferromagnetic transition with 
$T_{c}$ = 170 K.

Although several transition metals substitute for Ge in the type-I
clathrate structure,\cite{cordier91} attempts at Fe substitution 
have failed due to competing Fe-Ge phases.  However, we find 
that higher temperature synthesis produces a Fe-containing clathrate of the 
recently discovered chiral structure type, 
Ba$_{6}$Ge$_{25}$.\cite{fukuoka00,carrillo00,kim00} This clathrate 
contains an open Ge-Ge bonded framework, threaded by closed 
dodecahedral cages.
The material actually forms as Ba$_{6}$Ge$_{24}$, with vacancies 
occurring preferentially by a Zintl mechanism.\cite{kim00} 

\begin{figure}
\includegraphics{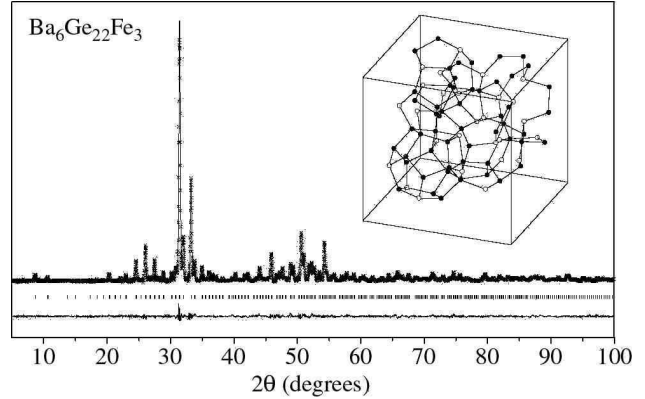}%
\caption{\label{fig:fig1}X-ray refinement for 
Ba$_{6}$Ge$_{22}$Fe$_{3}$.  Upper curve: data with fit. Lower curve:
difference plot.  Ticks show peaks indexed according to the chiral 
clathrate structure. Unit cell with clathrate framework is inset, 
showing a distorted dodecahedral cage at lower left. 3-bonded sites 
shown as open circles.}
\end{figure}

Our method followed several steps. 
Elemental mixtures, with excess Ba, were placed 
in BN crucibles, sealed under Ar, and pre-reacted by rf induction. 
The resulting mixtures included type-I clathrates and FeGe$_{2}$. 
Characterization was by x-ray diffraction (Bruker D8 Advance Powder, 
Cu $K\alpha$ radiation), and refinement using GSAS software.\cite{gsas}
To complete the reaction and eliminate FeGe$_{2}$, the mixtures were
arc-melted several times in a water-cooled copper crucible. 
The ingots were then sealed in quartz ampoules and 
annealed at 700 $^{\circ}$C for 100 hours. By this method we obtained 
single-phase or nearly single-phase Ba$_{6}$Ge$_{25-x}$Fe$_{x}$ with 
$x$ up to 3. However, furnace melting and slower cooling through the 
melting point produced a mixture of FeGe$_{2}$
and clathrate phases. An x-ray analysis for $x$ = 3 is shown in 
Fig.~\ref{fig:fig1}. All observed peaks were indexed according to the cubic 
space group P4$_{1}$32 (\#213), as previously reported for 
Ba$_{6}$Ge$_{25}$.\cite{fukuoka00,carrillo00,kim00} The lattice 
constant for $x$ = 3 is $a$ = 1.45520 nm, slightly 
larger than reported for Ba$_{6}$Ge$_{24}$ ($a$ = 1.45483 nm).\cite{kim00}

Energy-dispersive x-ray spectroscopy (EDS) and electron diffraction 
measurements (JEOL JEM-2010 electron microscope, 100 kV) for the $x$ = 3 sample 
showed that Fe
was indeed incorporated in the clathrate phase. 
These measurements also showed trace amounts of Ge and FeGe$_{2}$ to 
be present, but below the x-ray detectability limit (a few percent). 
Thus a minor fraction of the Fe atoms are present as FeGe$_{2}$. 
No other phases were observed.
Analysis gave an average clathrate composition
Ba$_{6}$Ge$_{21.9}$Fe$_{3.1}$ for the $x$ = 3 sample, a Ge:Fe 
ratio close to the starting composition. The Fe content per formula 
unit was 3.1 $\pm$ 1, with variations observed between 
crystallites, however the crystallites appeared locally homogeneous, 
with no apparent clustering.   

The chiral clathrate structure (Fig.~\ref{fig:fig1}) has Ba 
situated in open channels as well as closed cages. In the Ge$_{25}$ 
framework, two of the six distinct sites have three Ge-Ge bonds, 
the others having four.  By contrast, type I and II clathrates 
have only 4-bonded Ge sites in approximately tetrahedral bonding configurations, 
and closed cages. In our x-ray analysis we placed Fe
predominantly on three of the six sites, similar to the occupation 
parameters for In-doped Ge$_{25}$ clathrate,\cite{kim00} and the refinement gave a 
composition Ba$_{6}$Ge$_{21.8}$Fe$_{3.0}$, in excellent agreement with 
the EDS result (Ba$_{6}$Ge$_{21.9}$Fe$_{3.1}$). However, since the Fe 
and Ge x-ray structure factors are similar, we found that Fe could be 
redistributed among framework sites to give a similar composition and 
nearly identical goodness of fit ($R$ values: $R_{wp}$ = 
0.077 and $R_{p}$ = 0.053). The specific Fe locations therefore are uncertain,
although it is clear that they are distributed in the clathrate framework.

\begin{figure}
\includegraphics{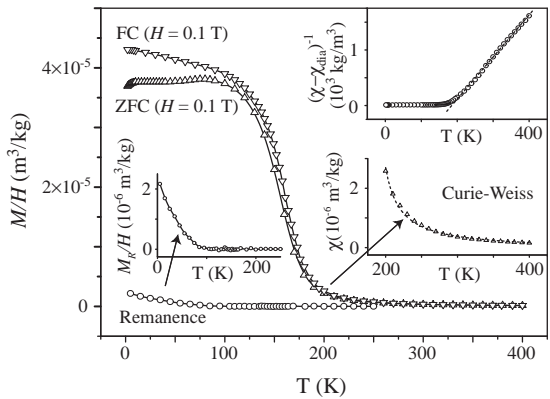}%
\caption{\label{fig:fig2}Temperature dependence of ZFC, FC and remanent 
magnetization of Ba$_{6}$Ge$_{22}$Fe$_{3}$ for $H$ = 0.1 T. Insets at right 
show FC magnetic behavior at high temperatures, with dashed curve 
from a Curie-Weiss fit. }
\end{figure}

\begin{figure}
\includegraphics{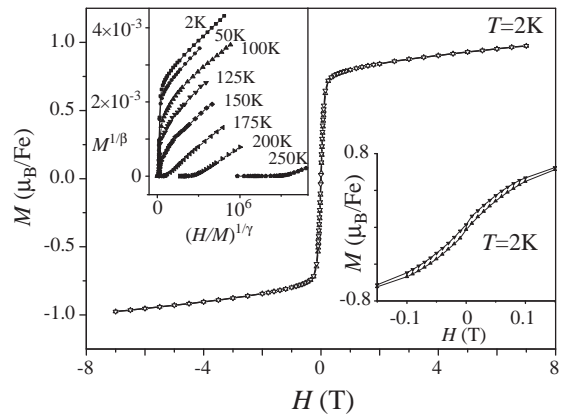}%
\caption{\label{fig:fig3}M-H hysteresis at 2 K, with 
the low-field region expanded on the right. At left is 
a modified Arrott plot of magnetization vs. field, with $\gamma$ = 
1.0 and $\beta$ = 0.45.}
\end{figure}

Magnetic measurements were made on a
Ba$_{6}$Ge$_{22}$Fe$_{3}$ sample using a SQUID magnetometer (Quantum Design MPMS-XL). 
Zero-field-cooled (ZFC) magnetization was measured on warming in 
fixed field, while field-cooled (FC) magnetization 
was measured vs. decreasing temperature in
constant field. Fig.~\ref{fig:fig2} shows data for 0.1 T applied field. 
The high-temperature behavior was fit to a
Curie-Weiss law, $\chi = \frac{C}{T-T_{c}} + \chi_{dia}$, where 
$\chi_{dia}$ denotes a diamagnetic background.
The value $C$ = 5.6 $\times 10^{-5}$ K m$^{3}$/kg indicates an effective moment 
p$_{eff}$ = 5.5 $\mu_{B}$ per Fe, close to the free-ion value, 5.9 $\mu_{B}$,
for high-spin Fe (p$_{eff}$ = $g[J(J+1)]^{1/2}$). $T_{c}$ obtained 
from the fit is 180 K. The diamagnetic term, 
$\chi_{dia}$ = $-8.2 \times 10^{-8}$ m$^{3}$/kg, 
is large and may be an overestimation due to rounding of 
the transition. Some distribution of $T_{c}$ seems likely because of 
the observed Fe concentration distribution.  However, low temperature 
magnetization measurements agree with the Curie fit; $T_{c}$ = 170 
K is estimated from a modified Arrott plot (inset to Fig.~\ref{fig:fig3}).

A divergence of $\chi_{FC}$ and 
$\chi_{ZFC}$ starts near 170 K, and becomes much more pronounced 
below 110 K (Fig.~\ref{fig:fig2}). The remanence was also
measured, warming the sample after removal of the field, and found to 
decrease with temperature, and vanish near 110 K. The magnetization at 2 K (Fig.~\ref{fig:fig3}) is polarized 
to 0.75 $\mu_{B}$ per Fe in a small field. However, full saturation 
is not reached even at 7 T, where the magnetization corresponds to 
1 $\mu_{B}$ per Fe.
This value is considerably lower than expected, given the value 
p$_{eff}$ = 5.5 $\mu_{B}$ obtained from the Curie fit.  This suggests a 
canted or noncollinear spin configuration.

The observed magnetic behavior cannot be due to the small amount of FeGe$_{2}$,
which has a spiral spin structure below 289 K, becoming commensurate at 
263 K.\cite{corliss85} These do not correspond to the observed 
transitions, and the observed moments clearly could not result from inclusions 
of this phase at the few percent level.

\begin{figure}
\includegraphics{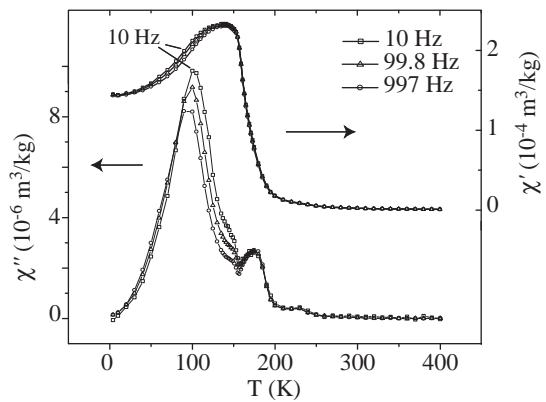}%
\caption{\label{fig:fig4}Ac susceptibility as function of temperature, measured
in zero dc field, with frequencies as shown.}
\end{figure}

To further understand the magnetic behavior, 
in-phase ($\chi'$) and out-of-phase ($\chi''$) ac susceptibilities
were measured under a 0.6 mT ac field, with the 
results shown in Fig.~\ref{fig:fig4}. The higher-temperature
data could be fit to a Curie 
law with $T_{c}$ = 180 K, in accord with the dc results. 
$\chi'$ exhibits a strong rise below 200 K, corresponding to the 
ferromagnetic transition. However, below a maximum, $\chi'$ 
decreases at lower temperatures and becomes frequency-dependent, while
$\chi''$ also becomes frequency-dependent.
The frequency dependence, combined with the decrease of 
ZFC vs. FC magnetization below 110 K, are characteristic of reentrant 
spin glass systems.\cite{binder86,mydosh93} We conclude that the transition near 
170 K is ferromagnetic, while the 110 K irreversibility point 
corresponds to a spin glass freezing transition. 

Frequency dependent $\chi_{AC}$ is found in spin glasses,
while for ordinary ferromagnets the time-scale is much too small, even for the 
critical slowing down region near 
$T_{c}$.\cite{binder86,mydosh93} Canted ferromagnets
with a low-temperature antiferromagnetism have exhibited ac behavior 
similar to that of a reentrant spin glass,\cite{mukherjee94} however $\chi''$ 
in that case shows only an increase with frequency, and no shift in
canting transition 
temperature.  The variation of $\chi''$ peak position,
$T_{p}$, per decade of frequency, $[dT_{p}/d(\log f)/T_{p}]$, is 0.042 for 
Ba$_{6}$Ge$_{22}$Fe$_{3}$, which is in the range of values 
observed for spin glasses, 0.004 to 0.08.\cite{mydosh93} 

Spin-glass reentrance is attributed to the freezing of transverse degrees of 
freedom in a ferromagnet containing disorder.\cite{gabay81} 
Alternatively, a cluster-glass mechanism has 
been argued,\cite{rakers87} and it is clear that ferromagnetic 
clusters and considerable disorder in the ferromagnetic 
state\cite{bao99} do play a role in real materials. However, a cluster glass 
alone does not explain the present data, particularly the 
low-temperature saturation. A 
remanence increase on cooling in the glassy state is also
typical of such spin glasses.\cite{wynn98} 

Fe occupancy of framework sites is below the percolation 
threshold, even for second-neighbor bonds, so it is likely that the 
mechanism for magnetic ordering is conduction-electron mediation, as in diluted magnetic 
semiconductors.  For comparison, in (Ga$_{1-x}$Mn$_{x}$)As, Mn carries about 5 
$\mu_{B}$, and $T_{c}$ scales linearly with small 
values of $x$, reaching $T_{c}$ = 110 K for $x$ = 0.05.\cite{ohno03}  
The present system is quite similar in $T_{c}$ vs. $x$. 
Band-structure calculations\cite{zerec02} indicate a gap below the 
Fermi level in Ba$_{6}$Ge$_{25}$ which may quite possibly be the E$_{F}$ 
location in the Zintl composition Ba$_{6}$Ge$_{24}$. The Zintl 
concept\cite{kauzlarich96} involves valence counting to give a 
closed-shell bonding configuration.  Assuming valence 3 for Fe, 
Ba$_{6}$Ge$_{21{3 \over 4}}$Fe$_{3}$ is also a Zintl phase, and very close to 
the average composition found here. Thus it seems likely that E$_{F}$ 
is near a band edge in this material, and the semiconductor 
model is appropriate.

There is considerable interest in the effects of disorder on the 
magnetic behavior of magnetic 
semiconductors;\cite{schliemann02,zarand02,dassarma03} the 
conduction-electron mediated interaction provides frustration, giving 
a non-collinear ordered state.\cite{schliemann02} For the case of 
(Ga,Mn)N, the ground state appears to be a spin glass with no 
ferromagnetism.\cite{dhar03}  Here we observe a 
ferromagnetic transition and spin-glass state in the same 
material.

In summary, a Ba$_{6}$Ge$_{22}$Fe$_{3}$ clathrate was 
synthesized, exhibiting a ferromagnetic $T_{c}$ = 170 K. 
Below 110 K, a transition to a reentrant spin glass was
evidenced by the onset of frequency-dependent ac susceptibility and a 
characteristic field-dependent magnetization.

\begin{acknowledgments}
This work was supported by the Robert A. Welch Foundation, 
Grant No. A-1526, and by the National Science Foundation (DMR-0103455).
\end{acknowledgments}

\bibliography{clathrateFe3}

\end{document}